**Near-field analysis of bright and dark modes on plasmonic metasurfaces showing extraordinary suppressed transmission**

*Sabine Dobmann\*, Arian Kriesch, Daniel Ploss, and Ulf Peschel*

Sabine Dobmann, Arian Kriesch, Daniel Ploss and Prof. Dr. Ulf Peschel
Institute of Optics, Information and Photonics and Erlangen Graduate School in Advanced Optical Technologies, Friedrich-Alexander-University Erlangen-Nuremberg (FAU), 91058 Erlangen, Germany

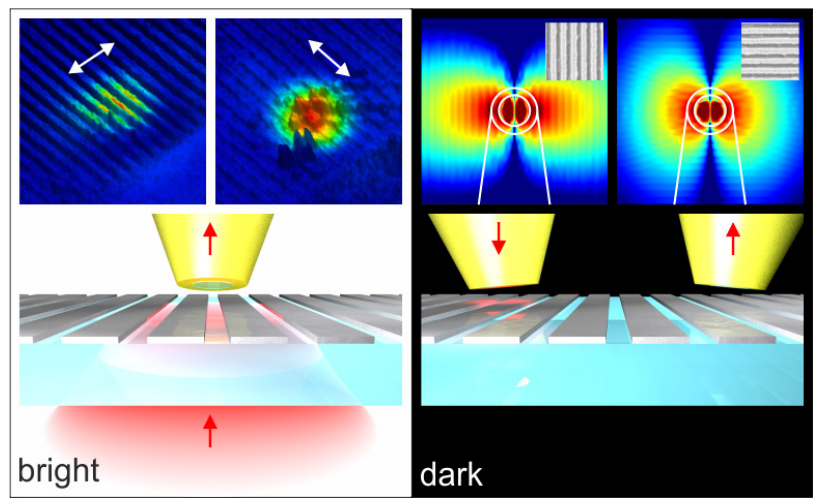

Plasmonic metasurfaces are investigated that consist of a sub-wavelength line pattern in an ultrathin (~ 10 nm) silver film, designed for extraordinarily suppressed transmission (EOST) in the visible spectral range. Measurements with a near-field scanning optical microscope (NSOM) demonstrate that far-field irradiation creates resonant excitations of antenna-like (bright) modes that are localized on the metal ridges. In contrast, bound (dark) surface plasmon polaritons (SPPs) launched from an NSOM tip propagate well across the metasurface, preferentially perpendicular to the grating lines.

# 1. Introduction

Artificial metamaterials[1] modify the flow of light in 3D[2]. Recently, their 2D relatives, so-called metasurfaces, have attracted particular interest[3]. Those metasurfaces usually consist, as in this letter, of structured thin metal films on a dielectric substrate. Metasurfaces made of materials with positive (dielectrics) and negative (metals) dielectric permittivity allow for the excitation of surface plasmon polaritons (SPP) and are promising candidates for various applications including negative refraction[4,5], phase shaping[6,7], polarization manipulation[8,9] and the design of compact color filters[10].

When free-space light impinges onto a metal film, it evanescently decays with a skin depth of only tens of nanometers. Hence, only extremely thin metal films transmit some radiation. This can be changed considerably by a sub-wavelength patterning resulting in so-called extraordinary transmission (EOT)[11–14], that is, an enhancement of the transmissivity far beyond the ratio between open areas and closed metal film[15,16].

Here we investigate metasurfaces that under illumination show the inverse physical phenomenon – so-called extraordinarily suppressed transmission (EOST)[17–19]. In this case, an ultrathin metal film that intrinsically transmits light is patterned to create antenna-like structures, which resonantly suppress transmission. This class of metasurfaces offers particular applicability and economic significance in optical nano-lithography, for example, by increasing contrast in reticles, while keeping them unprecedentedly thin[20].

Plasmonic metasurfaces modify the transmission in 3D[3,6] but may also support surface-bound plasmonic modes in 2D[21,22]. In both cases, nanoscale plasmonic resonances are excited that modify both loss and phase of the electromagnetic field. Understanding the interplay of antenna-like modes[23] and propagating bound SPP modes[24] is important to manipulate the interaction between light and metasurfaces.

Here we selectively excite different resonant modes and monitor them with a near-field scanning optical microscope (NSOM). We investigate a sub-wavelength patterned periodic plasmonic metamaterial that supports EOST in the visible wavelength range. Our structures consist of periodic arrays of lines with a pitch of p = 150 - 450 nm etched in a 10 nm ultrathin layer of Ag. We first analyze the far-field transmission (**Figure 1**a) and correlate the observed EOST with a characteristic field-enhancement, measured by NSOM (Figure 1b, f and g). By performing a double tip scan with an NSOM probe (Figure 1c) we furthermore demonstrate that this metamaterial modifies the in-plane propagation of guided plasmonic modes. Finally, we compare our measurements with supporting 3D finite difference time domain (FDTD) simulations. In particular the calculated band structure allows us to distinguish between fundamentally different modes, which are either excitable by the far-field (bright), thus causing EOST, or which are bound to the surface and therefore dark to the far-field. By comparing these different excitations we make a step towards a better understanding of optical modes supported by a metasurface.

## 2. Sample fabrication

The investigated metasurfaces consisting of an optimized metal array on a dielectric substrate were fabricated via magnetron sputtering and subsequent focused ion beam (FIB) milling. First we deposited a 10 nm thin Ag film on top of a 1 – 2 nm thin Ge adhesion layer[25] on a 170 µm thin BK7 glass substrate. After that, grating trenches were milled into the film using a FIB machine (Figure 1e). Multiple devices were fabricated covering a broad parameter range of different sub-wavelength pitches (p = 150 – 450 nm). Each grating field had a size between 5 µm x 5 µm to 20 µm x 20 µm. Samples were prepared in high quality with a typical line precision of 18 nm, which was only limited by the grain size (average ≤ 20 nm as determined with a scanning electron microscope (SEM)) of the metal films. The smallest feature size of the investigated structures was 60 nm (width of the metal trenches).

Full 3D finite difference time domain (FDTD) simulations were performed and indicated the presence of the EOST effect in the whole visible spectral range for a polarization perpendicular to the grating lines of an ultrathin, but structured metal film. EOST is the result of an enhancement of back reflection (Figure 1d) and a slight increase of absorption. Extensive parameter sweeps showed that the respective resonance position strongly depends on the width of the metal ridges but not on the pitch of the structure, thus excluding an effect based on grating diffraction. This supports the interpretation of EOST as a phenomenon based on the excitation of localized antenna-like resonances in the metal bars. This observation is in full accordance with more general theoretical investigations[23]. In contrast, extraordinarily enhanced transmission (EOT) as the inverse effect is commonly believed to be caused by propagating SPPs.[15] Hence, EOT and EOST do not seem to have the same origin.

**3. Far-field excitation of bright modes**

First, we determined the spectrally resolved reflection and transmission, the latter with additional variation of the angle of incidence, for various fabricated EOST samples with a far-field optical transmission probing setup (see supporting information for details).[26,27]

For the polarization perpendicular to the grating lines we find a broad transmission minimum accompanied by a maximum in reflection. In contrast, measurements with polarization parallel to the grating lines show an overall enhanced transmission due to the less dense metal film. The figure of merit to quantify the suppression of far-field transmission is $T_{rel} = (T_{Bulk} - T)/(T_{Bulk} + T)$. $T_{Bulk}$ represents the transmission of the unpatterned Ag film and $T$ is the transmission of the metasurface. Based on this definition, $T_{rel}$ can vary between the values -1 and +1 and represents the suppression in transmission independent of the original absolute transmission value (Figure 1d). The complete cancellation of the transmission of a structured Ag film is indicated by the value $T_{rel} = +1$.

The transmission of the unpatterned 10 nm Ag film was experimentally determined to be $T \approx 57\%$ ($\lambda = 633$ nm). This value is in good accordance to the theoretically predicted value (58 %). The maximum relative transmission for the measured 10 nm Ag thin metasurfaces was $T_{rel} \approx 0.88$, which corresponds to a transmission suppression of more than -12 dB compared to the transmission of the unpatterned film.

In order to further understand the underlying effect causing the cancellation of the transmission and redirection of the energy in the reflection channel, the angular dispersion[28] of the excited mode was investigated. The sample was mounted on a photodiode that collected the transmitted intensity; the air-gap between both elements was filled with index matching fluid. The sample was then rotated under a long working distance objective focusing the light on the structure (inset in **Figure 2**a). With this configuration the spectral transmission was probed for sample-axis angles from 0° to 60° in steps of 10° (Figure 2b). For comparison, we calculated the dispersion diagram of the investigated metasurfaces with a 2D finite difference frequency domain (FDFD) simulation in a broader parameter range that exceeds the range accessible in the experiment (Figure 2a).

The spectral position of the resonance stays nearly constant over the whole range of angles. The weak dependence on the angle of incidence is typical for antenna structures which do not significantly interact, but disagrees as expected with the phase-matching requirement for SPP excitation.

Geometrical parameter sweeps were the first test for this antenna model. For a constant thickness of 10 nm various width/pitch combinations were fabricated and the relative transmission $T_{rel}$ was experimentally determined for a wavelength of $\lambda_0 = 633$ nm. The underlying theoretical values were calculated with 2D FDTD simulations. The area of suppressed transmission in **Figure 3** (red) closely follows a hyperbola – indicating an EOST effect for a constant width of the metal ridges *w*. This is in good agreement with an interpretation of the effect as excitation and re-radiation of antennas formed by the metal ridges.

Hence, we suggest that the EOST is neither caused by a localized gap mode (Fabry-Pérot resonance) in the trenches nor by the excitation of planar SPPs but rather by an antenna-like stripe mode on the grating lines of the metasurface. However, from far-field measurements it is not possible to clearly distinguish which modes are actually excited and cause EOST and to what extent the mentioned effects involve the excitation of propagating SPPs.

In order to reveal the near-field distribution and to identify the electromagnetic modes[29,30], causing the extraordinary transmission effect, near-field measurements were performed while the sample was illuminated with focused, linearly polarized laser light. Near-field scans were recorded for both in-plane polarizations, parallel and perpendicular to the grating lines.

A linearly polarized beam was focused on the sample from the substrate side with a microscope objective numerical aperture of $NA = 0.85$ ($NA_{eff} = 0.4$, see supporting information). Its wavelength was chosen such that the EOST resonance of the structure was excited.

An aperture-NSOM[31] with a metal-coated tapered, bent optical fiber[32,33] operating in tapping mode scanned the modal near-field features of the laser spot impinging on the array (**Figure 4**a). Every fiber NSOM tip used in the experiments was made from multimode optical fiber. The tip apex was specifically cut and polished with the FIB to ensure a well-defined, reproducible diameter, shape and a planar scanning facet[34]. With the aid of the SEM we characterized the tapered, metal coated (Au/Cr) tips to have a sub-wavelength central silica aperture of ~ 240 nm and a total tip front facet diameter including the metal of ~ 800 nm. The selected tips had a small enough aperture to resolve the localized excitations on the grating lines.

The average tip oscillation amplitude was experimentally estimated as 30 nm for typical tips. Therefore, the tip sample distance is within the limits of the sample near-field. Evanescent near-field components from the metasurface were collected via the NSOM tip and guided by the connected fiber to an avalanche photodiode (APD). While scanning over the sample, the topography information and near-field intensity were acquired simultaneously.

As the grating period is deeply sub-wavelength, no propagating SPPs can be excited. Still, a strong interaction between light and metal is observed for an excitation normal to the grating lines. The experimental results (see Figure 4c) support the claim that EOST is mainly caused by localized excitations of the single building blocks of the grating structures. By superimposing the topographic scan with the optical signal (Figure 4c left) we prove that the excited mode is localized on the grating lines and not in the trenches. This shows that not a gap mode but a stripe mode is excited. For an excitation at resonance, such antenna-like[35] modes re-radiate with a phase shift of $\pi$ [36], thus causing destructive interference with the directly transmitted incident field, effectively suppressing any transmission.

We compare the NSOM measurement results with a 3D FDTD simulation (Figure 4b), where the grating is excited at resonance and with both polarizations. Similar to the experimental results the field is localized on the grating lines in case of a resonant excitation (polarization normal to the grating lines) resulting in a high field enhancement. Also, these simulations confirm the required $\pi$ phase shift of the excitation on the grating lines in comparison with the partly transmitted light in the trenches. Consequently, the far-field transmission is cancelled due to destructive interference between the emission of the excited ridges – acting as antennas –and the transmitted light of the sample structure. In case of an excitation with a polarization parallel to the lines the field interacts much less with the metal and does not resolve the grating structure (Figure 4b right), again in agreement with the experiment (Figure 4c right).

In all experimental scans and for all numerical simulations the size of the excited area never exceeds the focus spot size on glass. This is another indication for the strong localization of the resonantly excited modes on the metal bars. In conclusion, coupling between the grating lines is weak. For a resonant excitation via an incident beam we do not observe any other than strongly localized fields.

**4. Near-field excitation of dark modes**

Planar metal surfaces support propagating surface plasmon polariton (SPP) modes, which cannot be excited from the far-field. However, illumination with the near-field of an NSOM probe in close vicinity to the surface is a widely applied technique to excite and probe modes that intrinsically do not interact with the far-field. For a complete understanding of the plasmonic resonances that are supported by the EOST metasurfaces both radiative (bright) and non-radiative (dark) modes must be investigated.

To this end, we injected light with one NSOM (illumination) tip and scanned the surface with a second NSOM (collection) tip simultaneously[37] (double-tip scanning mode). All scans (**Figure 6**) showed good reproducibility even if the tip was retracted for sample alignment, subsequently approached and again positioned relative to the illumination tip. The closest distance of the two tips, determined via microscope imaging, was estimated to be below 500 nm. As very little light is detected in a double tip scan we now used tips with a larger diameter to probe the surface-bound modes with a reasonable signal-to-noise ratio. Although this caused a reduction of resolution we could still probe the near-field and in particular bound SPPs. The NSOM tips in use (aperture ø ≈ 1 μm) were measured to typically transmit $T_{tip} \approx 10^{-4}$, which is in accordance with other studies[34].

Consequently, the total transmission of our double tip scan as given by $T = T_{tip}^2 \, T_{sample}$ was extremely low. The power of the laser light that was filtered spectrally by an acousto-optical tunable filter (AOTF) and coupled into the optical fiber connected to the NSOM tip was limited to $P_{AOTF} \approx 2$ mW. For detection of the low light levels that came out of the scanning tip fiber, a single photon detector avalanche photodiode (APD) was used and operated at its detection sensitivity threshold.

We first characterized the input field by imaging the emitting tip (**Figure 5**a) with a microscope (Figure 5b) using different polarizers in the imaging beam path (Figure 5c and d). The polarization-resolved image was very stable and did not change in shape, even for various input conditions. The field emitted by the tip to the far-field was found to be dominantly linearly polarized perpendicular to the plane of the tapered fiber bend. In addition, a fraction of an orthogonally polarized field was detected, which can be attributed to a contribution from an azimuthally polarized mode (Figure 5f). These observations agree well with other recent investigations.[32] Only the linearly polarized mode develops strong longitudinal field components in the near-field at the edges of the aperture and can, therefore, excite SPP modes on the probed surface. Hence, in the subsequent discussion of the experimental results we will concentrate on the excitation with a linearly polarized tip mode.

Again we performed extensive numerical simulations to support our experimental investigations. We simulated the complete experimental configuration[32], including the excitation via the NSOM tip and the sample. The tapered apex of the fiber tip was modeled as a metallized and truncated cone (1 μm inner glass core at the tip front facet, continuous 250 nm Au coating, taper angle ~ 18°) with geometry parameters as determined from SEM images. In agreement with experimental observations discussed above the field in the NSOM tip was assumed to have the shape of a linearly polarized eigenmode of the metallized silica core with a polarization normal to the plane of the fiber bend (Figure 5e left). When leaving the plane-cut end facet of the tip (Figure 1f) the field spreads around the edge of the metal layer, thus developing strong vectorial components, which are perpendicular to the metal interface of the sample.

The resulting field distribution (Figure 5e right) resembles that of two electrical dipoles with opposite polarization that are oriented perpendicular to the sample surface. Those strong longitudinal components are not observed for completely metallized tips, but occur already for very small apertures and are very pronounced for the 1 μm diameter tips used in this study. The polarization of the two dipoles on the tip interface matches well the vectorial field of a surface plasmon polariton (SPP) propagating on the plane metal-dielectric interface[38,39]. When positioned close enough to the interface, each of the two dipoles excites propagating SPPs resulting in a characteristic two-lobe pattern (Figure 5g), where the power in the middle between both excitation sources is extinct due to destructive interference. This field pattern is in accordance with previous investigations of the field distribution of typical fiber NSOM tips[40–42] and to the recently reported pattern of excitation of SPPs via illuminated nanotubes[43]. Our NSOM scans on the plane Ag film show the predicted two-lobe structure (Figure 6e and i). However, in the experiment we do not observe a perfect zero, but a pronounced minimum because of possible asymmetries of the emitting tip and the finite size (aperture diameter of ~ 1.2 μm) of the collecting tip (Figure 6 c, e, g and i). Additional scans with a smaller collection tip (see supporting information) showed a more pronounced zero-node with the down-side of lower values of detected power, thus effectively limiting the extension of the field of measurement.

It was already generally predicted that an anisotropic metasurface as, for example, a grating will modify the propagation of SPPs compared with a flat metal film.[44,45] In order to launch light on the metasurface the input tip was placed at the boundary between the patterned and the plane metal film and the area in front of the input tip was scanned (Figure 6a and b) for two orientations of the grating lines. In all cases a reference scan on a piece of plane metal film close to the grating was taken. For a quantitative comparison we summed all the measured data along one direction, thus projecting all the detected power either on the vertical scan axis v (Figure 6d and h) or the horizontal scan axis u (Figure 6f and j). For all double-tip scans the diameter of the collection tip was larger than the small pitch of the grating. Therefore, the near-field scans resolve the field pattern of the propagating modes while most of the sub-wavelength pattern is averaged out.

In all measurements we observed the characteristic two-lobe pattern (see Figure 6), but its shape was considerably altered by the presence of the grating structure.

We first investigated a metasurface with an orientation of the grating lines parallel to the plane of the fiber bend and orthogonal to the polarization of the mode in the emitting tip (Figure 6c - f). Respective scan images show a more pronounced node of the two-lobe pattern (Figure 6f), which is in good agreement with our simulations. The propagation normal to the lines is favored and, therefore, the two-lobe pattern becomes elongated in the horizontal scan axis compared to the plane metal film (see Figure 6f). In contrast, the decay of the field in the vertical scan axis direction is not altered by the presence of the grating (Figure 6d).

When the stripes of the metamaterial are oriented perpendicular to the plane of the fiber bend and parallel to the polarization of the field in the emitting tip (Figure 6g - j), the excited SPPs propagate preferentially into the grating field (Figure 6g). Light is also slightly back-reflected at the boundaries of the patterned area causing a homogeneously illuminated background with the shape of the rectangular metasurface field. Also, the integrated scan images demonstrate that contrary to the previous case the two-lobe pattern is not altered parallel to the lines in horizontal scan axis direction u (Figure 6j), but extended perpendicular to the ridges in vertical scan axis direction v (Figure 6h) when compared with the plane metal reference scan.

To conclude, the propagation of guided modes is significantly affected by a nanostructure, but in a counter-intuitive way: Light propagation on a metallic metamaterial is considerably supported by the sub-wavelength voids in the structure. The SPP that propagates normal to the grating experiences reduced damping, due to its lower confinement to the lossy metal. We measured the effective propagation length for a $1/e$ decay of the intensity for both cases. The propagation length parallel to the lines coincides with that of the pure Ag (Figure 6d) whereas in case of propagation perpendicular to the lines (Figure 6h) a difference of 28 % was determined in comparison to the Ag film.

3D FDTD simulations agree very well with these experimental results (**Figure 7**). SPPs were excited inside the metasurface fields in both orientations (Figure 7a and c) and on a plane Ag film (Figure 7b). This allows us to directly compare the field evolution in the different domains.

In the simulation we followed the evolution of the electric field component $E_z$, which is normal to the metasurface plane and dominant in case of propagating SPPs. The monitor was placed 75 nm above the metal interface, which is assumed to be the mean height of both NSOM tips in the experiment. In agreement with our experimental observations, SPPs propagate further perpendicular to the array lines even compared to the propagation on the unpatterned metal film.

To better understand the observed phenomena, we analyzed the band structure of the periodically structured metasurface for excitations being homogenous along and polarized perpendicular to the lines.[28,46] We numerically calculated all supported modes of the system in a frequency range larger than the investigated one (**Figure 8**a), which allows us to directly compare and identify the experimentally measured modes.

"Bright" modes are placed in the light cone and therefore couple to the incident far-field, thus causing EOST but have low in-plane coupling. Based on sub-wavelength resolution NSOM scans we identify those modes as being localized on the metal ridges of the grating which act like optical antennas. We observed no significant in-plane propagation. In fact, the band diagram (Figure 8a) shows one localized antenna like SPP mode (1$^{st}$ LASPP) for $\lambda_0 \approx 680$ nm with a very low dispersion throughout the full 1$^{st}$ Brillouin zone, but particularly in the range of in-plane wave vectors $k_x \approx 0 - 8$ μm$^{-1}$ that we probed with the NA ≤ 0.9 of our far-field excitation. This is fully supported by the measured effectively zero-shift of the resonance wavelength over the variation of the angle of incidence, which is an experimental sweep of the in-plane k-vector (Figure 2b). This is characteristic for a localized SPP (LSPP) that does not propagate away from the point of excitation. The respective electric field distribution (Figure 8c) is localized at the metal ridges and has a low field overlap with adjacent modes, thus supporting the interpretation. It is interesting to note that additional localized modes exist: (1) a second-order antenna resonance LASPP (Figure 8b) at $\lambda_0 \approx 480$ nm, which is not excited in our far-field investigations and further higher order resonances (see supporting information), (2) a localized gap SPP resonance (LGSPP) (Figure 8e) at $\lambda_0 \approx 1$ μm that according to other calculations (Figure 2a) does not cause EOST.

In contrast, „dark" modes lie outside the light cone and cannot be excited from the far-field. But due to the increased coupling between single antennas they support the propagation of bound SPP modes inside the metasurface. We experimentally excited and probed these modes with an NSOM tip, directly coupling to the near-field. The dispersion relation of SPPs on a planar, unstructured Ag film with 10 nm thickness (Figure 8a dark red curve) branches off from the light line of the silica substrate (Figure 8a lower blue curve), finally converging towards a horizontal asymptote for increasing wave numbers. The periodic structuring of the metasurface modifies this dispersion relation, giving rise to the formation of a metasurface SPP (MSPP). This hybrid state results from the coupling between a conventional SPP and the localized antenna-like mode (1st LASPP). We observe the resulting anti-crossing close to the boundary of the 1st Brillouin zone at $k_x = 14.28$ μm$^{-1}$. The slope of the dispersion relation shows that such in-plane bound modes do not propagate ballistically, but spread diffusively in the 2D plane like massive particles (Figure 8d). Both resonances - bright and dark - are spectrally broad and therefore have an overlap with our excitation wavelength at 700 nm.

## 5. Conclusion

Concluding from our measurements and simulations we interpret the observed phenomena as based on the excitation of antenna resonances: once "bright" and singular, when far-field is transmitted through the metasurface in 3D and experiences EOST, once "dark" and collective, when modified metasurface SPPs (MSPPS) propagate along the metasurface. The propagation of the latter ones is considerably affected by the anisotropy of the metasurface.[44]

We have based our investigations on the simplest structure, a one-dimensional grating etched into an ultrathin Ag film. Similar effects are expected to occur for other plasmonic materials and more complicated structures.

**Experimental Section**

*Experimental Subheading*: Experimental Details. 12 point, double-spaced. References are superscripted and appear after the punctuation.[6]

*Numerical simulations* were implemented as full 3D finite difference time domain (FDTD) simulations with the commercial package FDTD Solutions by Lumerical Inc. Special emphasis was placed on the correct simulation of the various experimental settings.

*Fabrication* of the samples was realized with an AJA magnetron sputtering device and a Zeiss 30 kV Dual Beam Focused Ion Beam (FIB) machine with a Raith Elphy Plus pattern generator system. The working ion beams of 0.3 pA and 1.0 pA resulted in a minimum beam diameter of approximately 7 nm. The Dual Beam machine was also used for high-resolution scanning electron (SEM) microscopy and for taking cross-sections and geometric characterization of the samples. All processes were optimized for optimum grain-size and exact transfer of the designs.

*The light source* is a supercontinuum laser Koheras SuperK Extreme from Koheras/NKT Photonics. The light is filtered with an acousto-optical tunable filter (AOTF) that is also manufactured by NKT Photonics and delivers a $3 - 5$ nm line-width in the full visible wavelength range that we control with specifically made NI LabView-control software.

*Far-field transmission measurements* were undertaken using a specifically constructed free-space optical setup optimized for the visible wavelength range. The wavelength was scanned in steps of 10 nm over a spectral range of $\lambda = 470 - 720$ nm, which is sufficient to resolve the spectrally broad EOST response as determined by FDTD simulations. Precise adjustment is ensured and scans were taken with a Physik Instrumente (PI) xyz linear piezo stage with feedback and specifically made control software. The optical setup follows the design scheme as previously published[26,32,33], described in the text and the supplementary.

*Near-field scanning optical microscopy (NSOM)* measurements were taken with a custom modified dual probe NSOM system, originally manufactured as MultiView 4000 by Nanonics Imaging Ltd. The scanning probes were made from commercially available optical fiber that was pulled, tapered and coated with Cr/Au. We processed and characterized each single NSOM tip before we took the presented near-field scans with the methods described in [32]. The scanned light was detected with a Perkin Elmer APD SPCM-ARQRH single photon counting module with $\geq 70\%$ nominal photon detection quantum efficiency. All NSOM tips were post-processed and geometrically characterized with a Zeiss Dual Beam NVision 40 Focused Ion Beam (FIB) system.

**Supporting Information**
Supporting Information is available from the Wiley Online Library or from the author.

**Acknowledgements**
The authors thank P. Banzer, T. Bauer, A. Erdmann and H. Pfeifer for inspiring discussions paving the way for new ideas and E. Butzen for support with sample fabrication. This work was supported by the Cluster of Excellence Engineering of Advanced Materials (EAM), Erlangen. We acknowledge use of the facilities of the Max Planck Institute for the Science of Light (MPL), Erlangen. S.D., A.K. and D.P. also acknowledge funding from the Erlangen Graduate School in Advanced Optical Technologies (SAOT) by the German Research Foundation (DFG) in the framework of the German excellence initiative.

Author contributions: S.D., A.K. and U.P. devised the experiments. S.D. and D.P. optimized, fabricated and characterized the samples. S.D. conducted the numerical simulations. S.D. and A.K. performed the near-field investigations and A.K. proposed the interpretation scheme. All authors contributed to the final version of the manuscript.

# References

[1] N. I. Zheludev, Y. S. Kivshar, *Nat. Mater.* **2012**, *11*, 917.

[2] L. Verslegers, P. B. Catrysse, Z. Yu, J. S. White, E. S. Barnard, M. L. Brongersma, S. Fan, *Nano Lett.* **2009**, *9*, 235.

[3] A. V. Kildishev, A. Boltasseva, V. M. Shalaev, *Science* **2013**, *339*, 1232009.

[4] V. M. Shalaev, *Nat. Photonics* **2007**, *1*, 41.

[5] S. P. Burgos, R. de Waele, A. Polman, H. A. Atwater, *Nat. Mater.* **2010**, *9*, 407.

[6] F. Aieta, P. Genevet, N. Yu, M. A. Kats, Z. Gaburro, F. Capasso, *Nano Lett.* **2012**, *12*, 1702.

[7] P. Genevet, N. Yu, F. Aieta, J. Lin, M. A. Kats, R. Blanchard, M. O. Scully, Z. Gaburro, F. Capasso, *Appl. Phys. Lett.* **2012**, *100*, 013101.

[8] N. Yu, F. Aieta, P. Genevet, M. A. Kats, Z. Gaburro, F. Capasso, *Nano Lett.* **2012**, *12*, 6328.

[9] Y. Zhao, A. Alù, *Nano Lett.* **2013**, *13*, 1086.

[10] S. Yokogawa, S. P. Burgos, H. A. Atwater, *Nano Lett.* **2012**, *12*, 4349.

[11] T. W. Ebbesen, H. J. Lezec, H. F. Ghaemi, T. Thio, P. A. Wolff, *Nature* **1998**, *391*, 667.

[12] F. J. García de Abajo, *Rev. Mod. Phys.* **2007**, *79*, 1267.

[13] J. Bravo-Abad, A. Degiron, F. Przybilla, C. Genet, F. J. García-Vidal, L. Martín-Moreno, T. W. Ebbesen, *Nat. Phys.* **2006**, *2*, 120.

[14] W. L. Barnes, W. A. Murray, J. Dintinger, E. Devaux, T. W. Ebbesen, *Phys. Rev. Lett.* **2004**, *92*, 107401.

[15] C. Genet, T. W. Ebbesen, *Nature* **2007**, *445*, 39.

[16] F. J. Garcia-Vidal, L. Martin-Moreno, T. W. Ebbesen, L. Kuipers, *Rev. Mod. Phys.* **2010**, *82*, 729.

[17] J. Braun, B. Gompf, G. Kobiela, M. Dressel, *Phys. Rev. Lett.* **2009**, *103*, 203901.

[18] S. Xiao, J. Zhang, L. Peng, C. Jeppesen, R. Malureanu, A. Kristensen, N. A. Mortensen, *Appl. Phys. Lett.* **2010**, *97*, 071116.

[19] D. Reibold, F. Shao, A. Erdmann, U. Peschel, *Opt. Express* **2009**, *17*, 544.

[20] S. Dobmann, D. Shyroki, P. Banzer, A. Erdmann, U. Peschel, *Opt. Express* **2012**, *20*, 19928.

[21] I. I. Smolyaninov, Y.-J. Hung, C. C. Davis, *Science* **2007**, *315*, 1699.

[22] L. Yin, V. K. Vlasko-Vlasov, J. Pearson, J. M. Hiller, J. Hua, U. Welp, D. E. Brown, C. W. Kimball, *Nano Lett.* **2005**, *5*, 1399.

[23] G. D'Aguanno, N. Mattiucci, A. Alù, M. J. Bloemer, *Phys. Rev. B* **2011**, *83*, 035426.

[24] I. S. Spevak, A. Y. Nikitin, E. V. Bezuglyi, A. Levchenko, A. V. Kats, *Phys. Rev. B* **2009**, *79*, 161406.

[25] W. Chen, K. P. Chen, M. D. Thoreson, A. V. Kildishev, V. M. Shalaev, *Appl. Phys. Lett.* **2010**, *97*, 211107.

[26] P. Banzer, U. Peschel, S. Quabis, G. Leuchs, *Opt. Express* **2010**, *18*, 10905.

[27] T. Bauer, S. Orlov, U. Peschel, P. Banzer, G. Leuchs, *Nat. Photonics* **2014**, *8*, 23.

[28] A. Christ, S. G. Tikhodeev, N. A. Gippius, J. Kuhl, H. Giessen, *Phys. Rev. Lett.* **2003**, *91*, 183901.

[29] G. Ctistis, P. Patoka, X. Wang, K. Kempa, M. Giersig, *Nano Lett.* **2007**, *7*, 2926.

[30] T. Rindzevicius, Y. Alaverdyan, B. Sepulveda, T. Pakizeh, M. Käll, R. Hillenbrand, J. Aizpurua, F. J. García de Abajo, *J. Phys. Chem. C* **2007**, *111*, 1207.

[31] E. Betzig, J. K. Trautman, T. D. Harris, J. S. Weiner, R. L. Kostelak, *Science* **1991**, *251*, 1468.

[32] D. Ploss, A. Kriesch, H. Pfeifer, P. Banzer, U. Peschel, *Opt. Express* **2014**, 22, 13744-13754.

[33] A. Kriesch, S. P. Burgos, D. Ploss, H. Pfeifer, H. A. Atwater, U. Peschel, *Nano Lett.* **2013**, *13*, 4539.

[34] J. A. Veerman, A. M. Otter, L. Kuipers, N. F. van Hulst, *Appl. Phys. Lett.* **1998**, *72*, 3115.

[35] L. Novotny, N. van Hulst, *Nat. Photonics* **2011**, *5*, 83.

[36] P. Bharadwaj, B. Deutsch, L. Novotny, *Adv. Opt. Photonics* **2009**, *1*, 438.

[37] A. Kaneta, R. Fujimoto, T. Hashimoto, K. Nishimura, M. Funato, Y. Kawakami, *Rev. Sci. Instrum.* **2012**, *83*, 083709.

[38] J. A. Dionne, L. A. Sweatlock, H. A. Atwater, A. Polman, *Phys. Rev. B* **2005**, *72*, 075405.

[39] M. I. Stockman, *Opt. Express* **2011**, *19*, 22029.

[40] B. Hecht, H. Bielefeldt, L. Novotny, Y. Inouye, D. W. Pohl, *Phys. Rev. Lett.* **1996**, *77*, 1889.

[41] L. Neumann, Y. Pang, A. Houyou, M. L. Juan, R. Gordon, N. F. van Hulst, *Nano Lett.* **2011**, *11*, 355.

[42] L. Novotny, D. W. Pohl, B. Hecht, *Ultramicroscopy* **1995**, *61*, 1.

[43] N. Hartmann, G. Piredda, J. Berthelot, G. C. des Francs, A. Bouhelier, A. Hartschuh, *Nano Lett.* **2012**, *12*, 177.

[44] Y. Liu, X. Zhang, *Appl. Phys. Lett.* **2013**, *103*, 141101.

[45] B. Stein, J.-Y. Laluet, E. Devaux, C. Genet, T. W. Ebbesen, *Phys. Rev. Lett.* **2010**, *105*, 266804.

[46] P.-C. Li, Y. Zhao, A. Alù, E. T. Yu, *Appl. Phys. Lett.* **2011**, *99*, 221106.

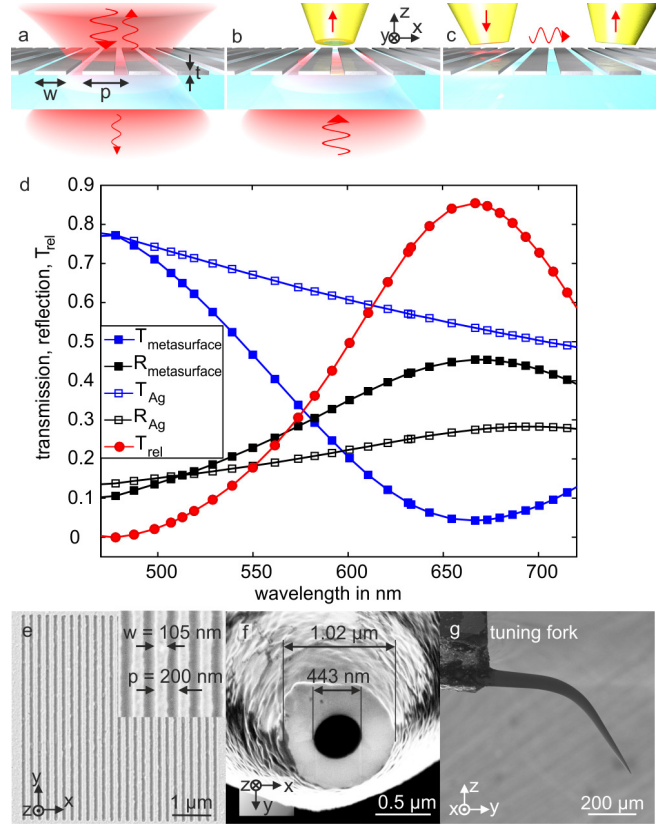

**Figure 1.** (a) The transmission and reflection of the metasurface grating are probed with a focused Gaussian beam to quantify the spectral EOST. The metasurface grating has the linewidth $w$, array pitch $p$ and thickness $t = 10$ nm. (b) It is excited from below through the substrate while NSOM scans detect the near-field distribution on the top of the sample. (c) The left NSOM tip evanescently excites a bound SPP that propagates along the metasurface and is simultaneously probed with a second NSOM tip. (d) Far-field transmission and reflection measurement results (as shown in (a), x-polarization) of a sample structure that exhibits EOST in comparison to those of an unstructured Ag film. The figure of merit for the suppressed transmission is $T_{\text{rel}} = (T_{\text{Bulk}} - T)/(T_{\text{Bulk}} + T)$ with an EOST resonance position at 667 nm. (e) SEM of the 10 nm thin metasurface characterized in (d). (f) Front facet of a cut and polished NSOM aperture fiber tip. (g) SEM of an NSOM tip mounted on the tuning fork.

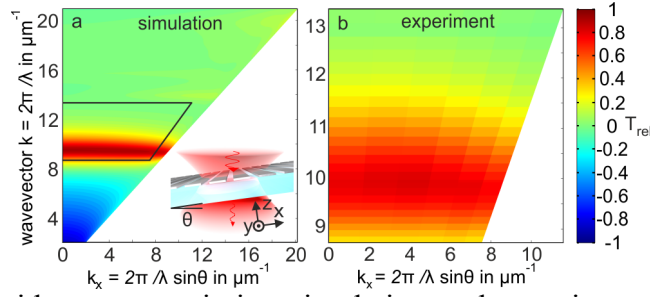

**Figure 2.** Oblique incidence transmission simulation and experimental results for a 10 nm thin metasurface structure ($w = 90$ nm, $p = 170$ nm, EOST resonance wavelength $\lambda_{res} = 630$ nm) with resonant polarization perpendicular to the grating lines. (a) Finite difference frequency domain simulation of angular dispersion of $T_{rel}$. The marked area indicates the experimentally accessible wavelength range. The inset shows the corresponding experimental measurement. By tilting the sample with respect to the static focus the angle of incidence θ is varied from 0° to 60° in 10° steps. This approach results in an in-plane $k_x$ vector sweep. (b) Measurement result of the angular dispersion of $T_{rel}$.

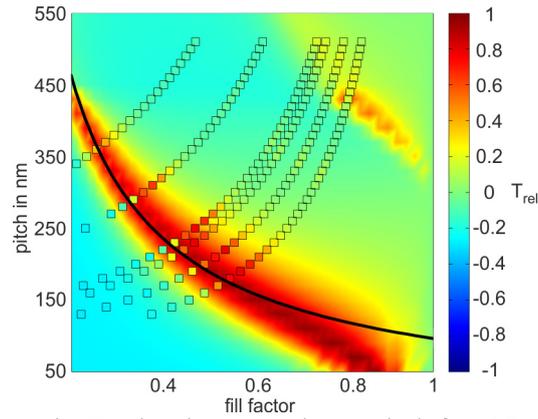

**Figure 3.** The figure of merit $T_{rel}$ is shown color-coded for 10 nm thin metasurfaces with varying fill factor and pitch combinations for a wavelength of 633 nm. The color plot shows the results of FDTD simulations and the superimposed squares indicate measured values of $T_{rel}$ (same color code) for metasurfaces with the respective lateral geometrical parameters (pitch precision ~ 10 nm, typical line precision ~ 18 nm). The hyperbola (black line) highlights a constant linewidth $w$ in the respective gratings.

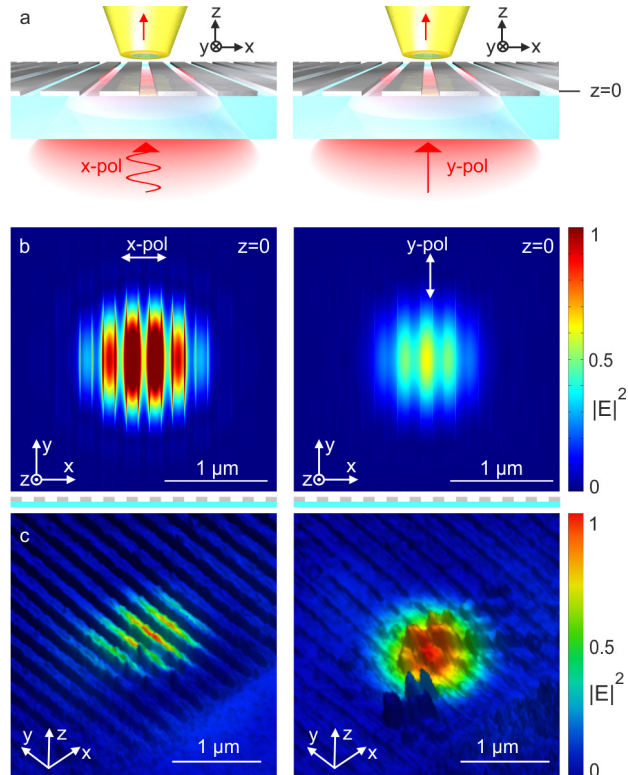

**Figure 4.** Near-field scans of metasurface modes that are excited at the EOST resonance wavelength ($w = 110$ nm, $p = 225$ nm). (a) Far-field light excites the metasurface grating with linear polarization either perpendicular (left) or parallel (right) to the array lines. (b) Simulation results show the electric field intensity in the plane $z = 0$ that corresponds to the upper interface of the metasurface. The positions of the grating lines and trenches are indicated below the plot. Both plots share the same color scaling. (c) 3D overlay of the near-field intensity of the NSOM scan on the measured topography. The intensities for both polarization directions are normalized to their individual maximum. (b) and (c) clearly show that for the resonant polarization (left) the electric field is localized on the metal ridges that act as antennas, while for the non-resonant polarization (right) the field is transmitted through the gaps of the grating.

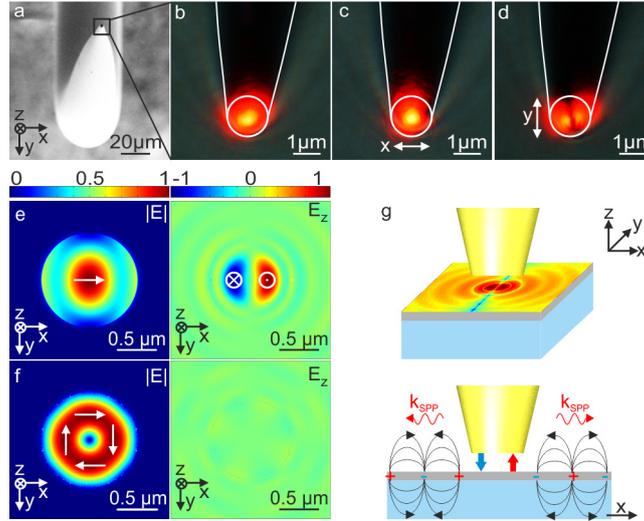

**Figure 5.** (a) SEM of a tapered and bent NSOM fiber tip with the apex plane (highlighted). (b-d) Microscope images of the tip emission with (b) no polarizer, (c) polarizer oriented perpendicular to the plane of the fiber bend and (d) polarizer oriented parallel to the plane of the fiber bend. (e,f) Calculated modes with lowest damping in the truncated cone forming the fiber taper. (e) Left: Mode profile of the linearly polarized mode in the tip plane 100 nm before the tip aperture. Right: Z-component of the electric field in plane of tip front facet shows two antiparallel z-dipoles (f) Left: Mode profile of the azimuthally polarized mode in the tip plane 100 nm before the tip aperture. Right: The z-component of the electric field in plane of tip apex is negligibly small. (g) Schematic demonstration of the evanescent near-field distribution and resulting SPP excited on the Ag film. The upper panel shows the resulting two-lobe pattern including the central node parallel to the plane of the fiber bend in case of excitation with two antiparallel z-dipoles from the linearly polarized fiber mode shown in (e). The lower panel shows the evanescent SPP coupling on the planar metal film from the antiparallel z-dipoles due to high evanescent mode field overlap.

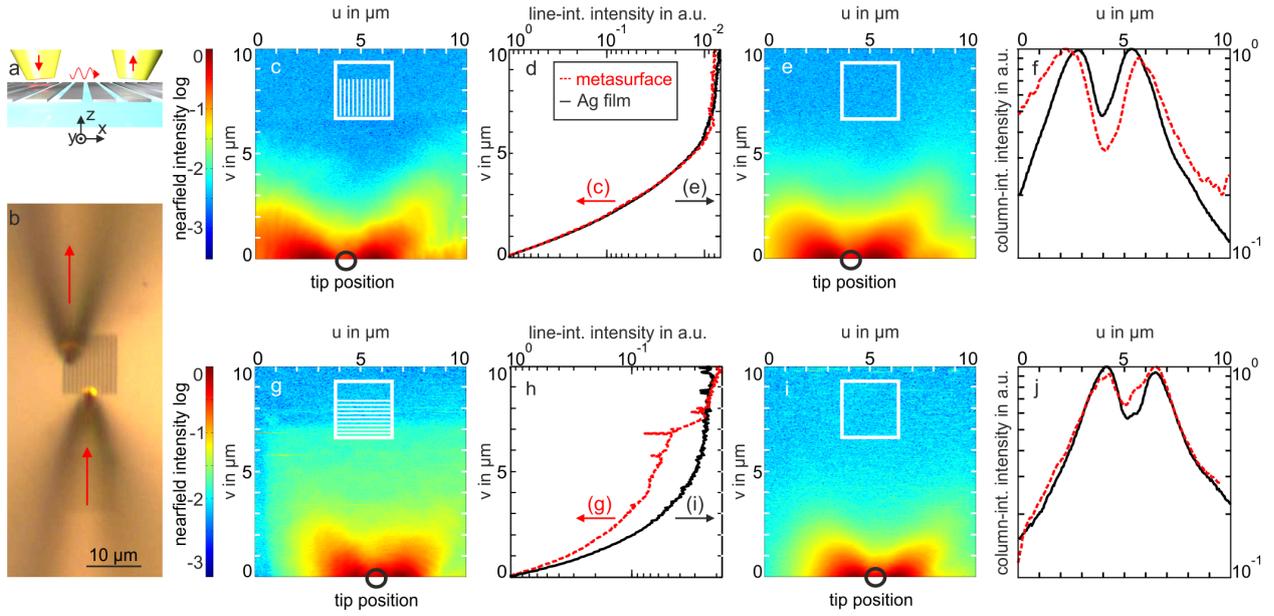

**Figure 6.** Double tip NSOM scans of a metasurface ($w = 120$ nm, $p = 220$ nm, $\lambda=700$nm, EOST resonance). (a) Basic principle and (b) microscope image of the two tips approached to the metasurface. (c,e,g,i) Scans of vertically (c) and horizontally (g) oriented gratings, compared with scans of plane Ag films (e,i). (d,j) Near-field intensities integrated perpendicularly to the grating lines almost coincide for the metasurface and the Ag film, showing that propagation parallel to the grating lines is not affected. (f,h) In contrast, integration along the lines results in large differences thus demonstrating that propagation length perpendicular to the grating lines is increased by the presence of the metasurface.

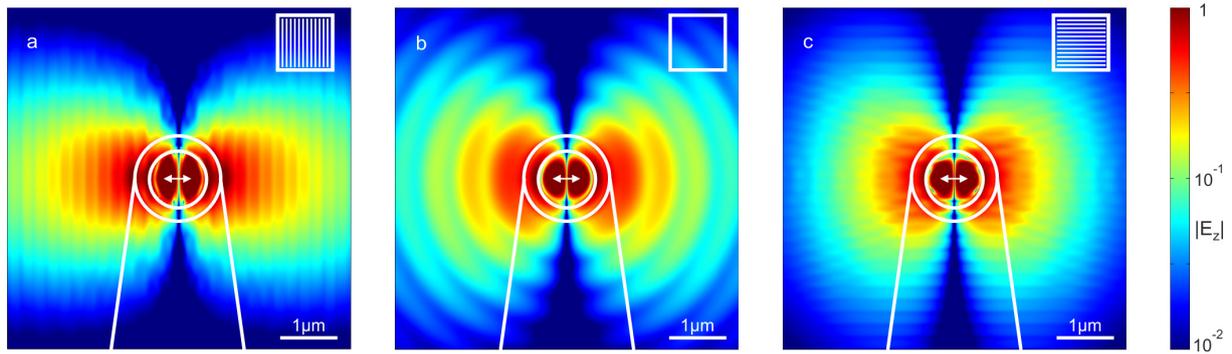

**Figure 7.** 3D finite-difference time-domain (FDTD) simulations demonstrating the propagation of SPPs excited by the near-field of the NSOM tip (Figure 5e) on (a) an anisotropic metasurface with lines oriented vertically and, therefore, parallel to the plane of the fiber bend, (b) a plane Ag film, (c) an anisotropic metasurface with lines oriented horizontally and, hence, perpendicular to the plane of the fiber bend. The main component of the electric field of propagating SPPs $|E_z|$ is plotted 75 nm above the surface, the estimated height for both tips in the experiment and simulation. In all plots the bold overlaid curves indicate the illumination tip position and orientation. All plots are scaled to the same value leaving some parts of the scan overexposed.

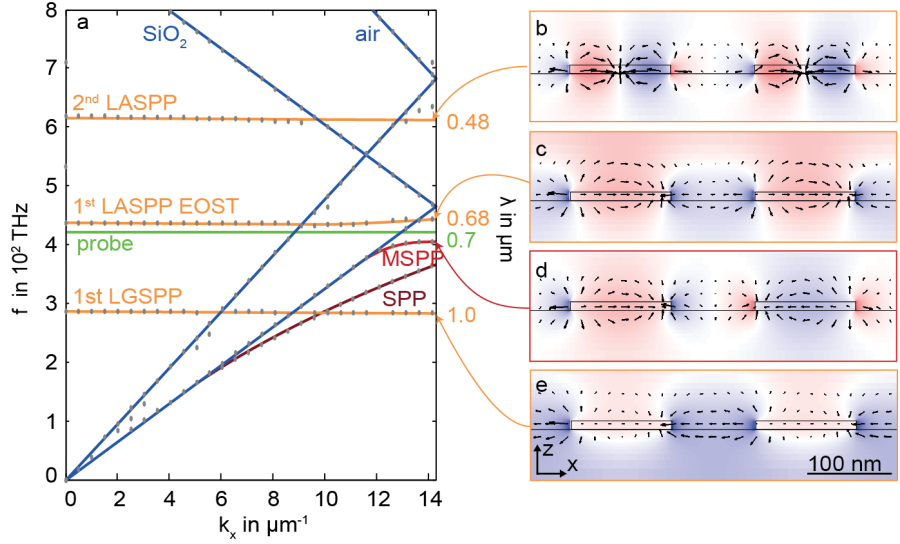

**Figure 8.** (a) Calculated modal dispersion (FDTD) of the investigated EOST grating structure ($w$ = 120 nm, $p$ = 220 nm, $t$ = 10 nm) for the first Brillouin zone. Only the mode maxima are depicted. The SPP mode of the unstructured 10 nm Ag film is included for comparison. (b-e) Respective mode profiles in a double pitch geometry for the indicated wavelength. The colorplot shows the $E_x$ component of the electric field, whereas the arrows represent the electric field orientation in the xz plane. (c) Dipolar 1$^{st}$LASPP responsible for EOST effect of the metasurface, (b) higher order localized antenna surface plasmon polariton (2$^{nd}$ LASPP), which can not be excited from the far-field, (d) propagating metasurface SPP close to the edge of the first Brillouin zone which also can only be excited from the near-field and (e) first localized gap surface plasmon polariton (LGSPP). In (b,d) the dark modes are shown whereas in (c,e) bright modes of the metasurface are presented.

Supporting Information

# Near-field analysis of bright and dark modes on plasmonic metasurfaces showing extraordinary suppressed transmission

*Sabine Dobmann\*, Arian Kriesch, Daniel Ploss, and Ulf Peschel*

**Numerical simulations**
Numerical simulations were implemented as 2D and full 3D finite difference time domain (FDTD) simulations with the commercial package FDTD Solutions by Lumerical Inc. Special emphasis was paid to the correct simulation of the various experimental settings. The complex material parameters were taken from tabulated literature data (1).

**Experimental setup for the far-field investigations**
Transmission and reflection measurements for the optical characterization of the extraordinarily suppressed transmission (EOST) samples were conducted with a custom-made optical setup (Figure S1). A collimated broadband laser beam from a super-continuum light source (NKT Photonics/Koheras, SuperK Extreme) was spectrally filtered in the visible wavelength range of 470 – 720 nm by an acousto-optical tunable filter (AOTF, NKT Photonics). The resulting monochromatic (bandwidth: 3 - 5 nm in the visible spectral range) beam is polarization filtered (extinction ratio 1:100,000) and subsequently impinges on a non-polarizing beam splitter (NPBS), where 50 % of the incident power is separated from the main path and detected by a Si reference photodiode. The beam then enters a high numerical aperture objective (NA = 0.9) that is aberration corrected for the visible light spectrum. For the used beam diameter the effective NA is 0.28 resulting in a focus spot size on the sample surface of ø ≈ 1 µm. Using this lower effective NA avoids conversion of field components as would happen with a highly focused beam and allows for direct comparison with plane wave simulations. The focused light impinges on the sample structure, which is positioned with a piezo scanning system (Physik Instrumente, PI, 5 nm xyz precision). The minimum 5 µm x 5 µm metasurface fields are laterally scanned through the static focus in steps of 250 nm. This approach allows for easier positioning and minimizes the influence of local inhomogeneities of the periodic structure. Below the sample structure in confocal position with the focusing objective a collecting objective (immersion oil, NA = 1.3) is mounted, which collects the light that is then guided to a Si photodiode to measure the sample transmission. The reflected light is gathered by the focusing objective and after passing the NPBS, the remaining 50 % are redirected on a Si photodiode to detect the sample reflection. Reflection and transmission signals were therefore measured simultaneously. The absolute values of transmission and reflection were determined by normalization with quasi-analytically calculated values for the glass-air interface. Power fluctuations due to the probing laser were eliminated by an additional normalization through the reference photodiode signal. The wavelength was scanned in steps of 10 nm over a spectral range of λ = 470 - 720 nm, which is sufficient to resolve the spectrally broad EOST response as determined by FDTD simulations.

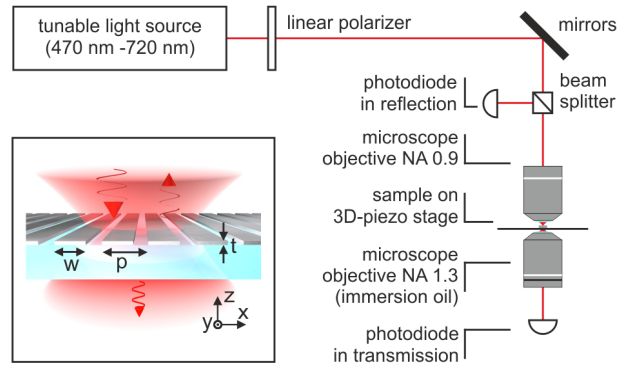

**Figure S1.** The confocal far-field scanning setup. Transmission and reflection signals of the metasurfaces were detected simultaneously. The inset illustrates the metasurface structure design with variable geometrical parameters (thickness of the Ag film $t$, width of the Ag ridges $w$, pitch of the periodic structure $p$).

In order to probe the angular dispersion of the metasurfaces the setup was modified. The focusing objective was replaced by a long working distance objective (Mitutoyo, NA = 0.4). The increase in working distance gave enough spatial freedom to rotate the sample below the objective. In this case, the sample was directly placed on top of a photodiode and the gap was filled with immersion oil.

**Sample fabrication**
The metal films were fabricated with an AJA magnetron sputtering device on borosilicate glass (BK7, thickness 170 µm) substrates. The maximum roughness of the substrate was specified as 3 nm. A 1 – 2 nm thick Ge adhesion layer was applied below the 10 nm thick Ag layer. The thicknesses of the Ge and Ag layers were extrapolated from focused ion beam (FIB) characterization measurements of thicker reference layers with the same machine and fabrication parameters.
The samples were structured with a Zeiss 30 kV dual beam focused ion beam (FIB) machine with an integrated nano patterning and visualization engine (NPVE). The applied ion beams with a current of 0.3 pA and 1.0 pA resulted in a minimum beam diameter of approximately 7 nm. The Dual Beam machine was also used for high-resolution scanning electron (SEM) microscopy and for taking cross-sections and for the geometric characterization of the samples. All processes were optimized for optimum grain-size and exact transfer of the designs.

**Experimental setup for the near-field investigations**
The near-field scanning optical microscope (NSOM) measurements were taken with a custom modified dual-probe NSOM system based on a commercially available fiber aperture NSOM (Nanonics Imaging Ltd., MV 4000). The scanning probes were made from commercially available multimode optical fiber that was pulled, tapered and coated with Cr/Au. The FIB processed fiber aperture tip was operated in tapping mode and the sample-tip distance was phase-controlled by atomic force feedback. The tips were raster-scanned across the sample with a lateral resolution of 40 nm while the tapered tip was collecting the near-field intensity present on the sample surface, which was then converted into propagating modes guided by the connected optical fiber to a high sensitivity avalanche photodiode APD (Perkin Elmer SPCM-ARQRH) single photon counting module with $\geq 70\,\%$ nominal photon detection quantum efficiency.

For simultaneous far-field excitation and near-field detection, the polarization of the focused light was prepared and aligned with the following setup.

The light of the super-continuum laser source (Koheras, SuperK Extreme) is spectrally filtered by the acousto-optical tunable filter (AOTF) and afterwards coupled into single-mode optical fiber that filters the optical mode and guides the light to the experimental setup. The light out of this fiber is collimated and guided through a $\lambda/2$ plate for power maximization and a polarizer before it is split by a NPBS. The reference path at the second output of the NPBS is used for power normalization. The remaining beam is guided to the center of the focusing objective below the sample structure.

The system is equipped with two xyz-piezo scanning towers allowing for a near-field illumination and simultaneous near-field detection measurement option. The two tips as well as the sample itself are mounted on in total three piezo scanner systems allowing for accurate positioning and scanning resolutions in the nm range.

**Preparation of near-field tips**

The commercial multimode fiber NSOM tips used in the experiments were fabricated by Nanonics Inc. for the Nanonics Multiview MW 4000 NSOM. They were mounted onto a geometrically adapted SEM/FIB Zeiss Dual Beam NVision 40 holder that aligns the front facet plane of the tip parallel to the $Ga^+$ ion beam axis of the FIB column. With this approach the tapered tip is cut parallel to the tuning fork mount and, therefore, parallel to the scanning surface. In a first step, a FIB beam current of 10 pA is used to open the tip to a defined diameter, which can be visually controlled by the SEM imaging system. Finally, a polishing step with 1 pA FIB current is used to finalize the procedure.

**High-resolution near-field scans**

In addition to the NSOM scans in the letter (Figure 6, collection tip diameter ~ 1.2 µm) where the propagation of modes on the metasurface was investigated with as low as possible additional loss, we conducted additional NSOM scans with a collection tip diameter of ~ 200 nm (Figure S2). In those scans, the node in the center of the two antiparallel dipoles is more pronounced (Figure S2a and b). With this smaller tip diameter and consequently higher optical resolution the excitation of the individual ridges is clearly visible in the scan (Figure S2a). The intensity variation coincides with the period of the metamaterial structure (Figure S2c). The two-lobe emission pattern is stable concerning various emission tip diameters in theory and in experiment. This clearly indicates that the emission by the tip can be well approximated by two anti-parallel dipoles with an orientation normal to the sample surface. The often cited (2) one-z-dipole emission pattern is only valid for closed tips.

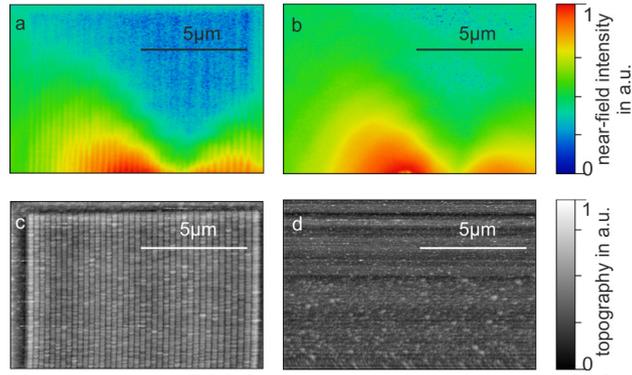

**Figure S2.** Emission pattern of a NSOM tip scanned with a ø = 200 nm aperture NSOM tip (a) on a metasurface and (b) on a pure Ag film. (c,d) Topographic scan corresponding to optical signal in (a,b), respectively.

**Band diagrams**

The dispersion relation of the modes on the unstructured metal film (Figure S3a) and the band structure of the investigated metasurface (Figure S3b) with a geometry corresponding to Figure 6 of the letter were calculated with FDTD solutions by Lumerical Inc. For the plain metal film, the dispersion of the propagating surface plasmon polariton (SPP) (asymmetric coupled SPP mode on the thin metal film) is folding up on the edge of the first Brillouin zone. For the metasurface the various localized modes are visible as well as the propagating metasurface SPP (MSPP). Figure 8 of the letter was generated by a peak-search algorithm that searched the local field maxima in the frequency-k-space for the whole band structure, which is shown in Figure S3.

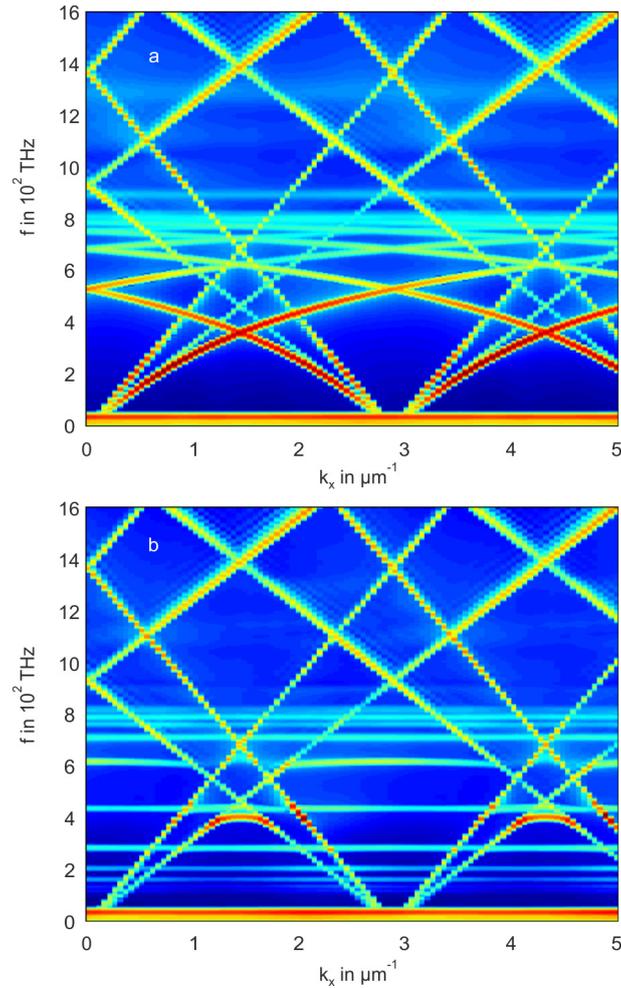

**Figure S3**. Band diagrams for (a) an unpatterned 10 nm thin Ag film and (b) the 10 nm thin investigated metasurface with $w = 120$ nm and $p = 200$ nm. The total electromagnetic field in frequency domain is plotted logarithmically.

**References**

(1) Edward D. Palik, *Handbook of Optical Constants of Solids*. Academic Press: Boston, **1985**
(2) Novotny, L.; Pohl, D. W.; Hecht, B. *Ultramicroscopy* **1995**, *61*, 1–9.